\documentclass[aps,prl,twocolumn,superscriptaddress,showpacs,letter]{revtex4}
\usepackage{amsfonts,amssymb,amsmath}
\usepackage{color}
\usepackage{graphicx}
\usepackage{dcolumn}
\usepackage{bm}

\begin{document}
\title{Projected p-wave superconducting wave-functions for topological orders}

\author{Su-Peng Kou}
  \email{spkou@bnu.edu.cn}
  \affiliation{Department of Physics, Beijing Normal University, Beijing 100875, China}

\author{Xiao-Gang Wen}
  \affiliation{Department of Physics, Massachusetts Institute of Technology, Cambridge, Massachusetts 02139}

\begin{abstract}

In this paper we develop a sysmatical theory for topological orders
by the projected $p$-wave superconducting (SC) wave-functions and
unify the different topological orders for spin models into the
fermionic picture. We found that
the energy for the fermions at $\mathbf{k}=(0,0)$, $(0,\pi )$, $(\pi ,0)$, $%
(\pi ,\pi )$\emph{\ }acts as a topological invariable to
characterize 16 universal classes of different topological orders
for the spin models with translation invariance. Based on the
projected $p$-wave SC wave-functions the topological properties for
the known topological orders in the exact solved spin models are
obtained. Finally new types of topological orders are predicted.


\end{abstract}

 \maketitle

Topological order is a new paradigm that lies beyond the realm of Landau's
theory based on the concept of symmetry-breaking for classical statistical
systems. Physically, the topological orders describe the quantum orders with
the long range entanglements in a gapped quantum ground state. A well-known
example for the topological order is the fractional quantum Hall (FQH)
state, which carries fractional charge and anyon statistics. To learn the
nature for the FQH states, Laughlin wave function is a good staring point.
Recently, many spin models with different topologically ordered states were
found, such as the toric-code model \cite{k1}, the Wen's plaquette model
\cite{wen,wen1} and the Kitaev model on a hexagonal lattice \cite{k2}. It is
interesting that all the topological orders of the exact solved spin models
can be described by $p$-wave superconducting (SC)\ wave functions \cite
{cn,wen1,yu,ki,zhou}. People may ask two questions : if the topological
orders for the different spin models are the same one? and how to use $p$%
-wave SC wave functions to describe the topological properties of the spin
models? In this paper, we would like to develop a sysmatical theory for the
topological ordered states based on the (projected) $p$-wave SC
wave-functions and learn the nature of the different kinds of topological
orders for the spin models.

\emph{The general ``mean-field'' spinless fermion Hamiltonian} - We start
with the general ``mean-field'' spinless fermion Hamiltonian with nonzero $p$%
-wave SC order parameter
\begin{equation}
H_{mean}=\sum_{ij}\psi _i^{\dag }u_{ij}\psi _j+\sum_{ij}(\psi _i^{\dag }\eta
_{ij}\psi _j^{\dag }+h.c.)+\mu \sum_i\psi _i^{\dag }\psi _i.  \label{ks}
\end{equation}
where $u_{ij}$, $\eta _{ij}$ are 2 by 2 complex matrices. $\mu $ is the
chemical potential. Let $|\Psi _{mean}^{(u_{ij},\eta _{ij})}\rangle $ be the
ground state of $H_{mean}$. To represent the topological order with full
gapped excitations, each site must have just single fermion. Then we add a
constraint to the $p$-wave SC wave functions. The topological ordered state
can be obtained from the mean-field SC state $|\Psi _{mean}^{(u_{ij},\eta
_{ij})}\rangle $ by projection it into the subspace with single fermion per
site, $|\Psi _{spin}^{(u_{ij},\eta _{ij})}\rangle =P|\Psi
_{mean}^{(u_{ij},\eta _{ij})}\rangle $. Here one has the projection operator
as $P=\prod_i\frac{1-(-1)^{\psi _i^{\dag }\psi _i}}2.$ \emph{Because the
fermion number can only be }$0$\emph{\ or }$1$\emph{\ on each site,} \emph{%
the constraint for the fermion number on each site can be reduced into a
constraint for total fermion number}! Because the fermion number for the
superconducting state is not conserved, we can only distinguish the ground
state with an even total fermion number or an odd one. On the even-by-even
(e*e), even-by-odd (e*o) and odd-by-even (o*e) lattices, the total fermion
number $N$ must be even, the projection operator turns into $P=P_r^e=\frac{%
1+(-1)^{\hat{N}}}2.$ On an odd-by-odd (o*o) lattice, the total fermion
number must be odd, the projection operator $P$ is reduced into $P_r^o=\frac{%
1-(-1)^{\hat{N}}}2.$ Here $\hat{N}=\sum\limits_i\psi _i^{\dag }\psi
_i=\sum\limits_{\mathbf{k}}\psi _{\mathbf{k}}^{\dag }\psi _{\mathbf{k}}$ is
the total fermion number operator.

For the translation invariant ansatz, one can rewrite the above
``mean-field'' fermion Hamiltonian in momentum space as $H_{mean}=\sum_{%
\mathbf{k}}\Psi _{\mathbf{k}}^{\dag }U(\mathbf{k})\Psi _{\mathbf{k}}\label{h}
$ with $\Psi _{\mathbf{k}}=\left(
\begin{array}{l}
\psi _{\mathbf{k}} \\
\psi _{-\mathbf{k}}^{\dag }
\end{array}
\right) $ and $\Psi _{\mathbf{k}}^{\dag }=\left(
\begin{array}{ll}
\psi _{\mathbf{k}}^{\dag } & \psi _{-\mathbf{k}}
\end{array}
\right) .$ We note that $({\psi }_{-\mathbf{k}}^{\dag },{\psi }_{-\mathbf{k}%
})$ can expressed in term of $({\psi }_{\mathbf{k}}^{\dag },{\psi }_{\mathbf{%
k}})$ from $\Psi _{-\mathbf{k}}=\Gamma \Psi _{\mathbf{k}}^{*}\ $and $\Psi _{-%
\mathbf{k}}^{\dag }=\Psi _{\mathbf{k}}^T\Gamma $ with $\Gamma =\sigma _1.$
Then we get that $U(\mathbf{k})$ has an additional constraint $U(\mathbf{k}%
)=-\Gamma U^T(\mathbf{k})\Gamma .\label{c}$ Under the constraint, we divide $%
4$ Emit $2\times 2$ matrices $M_\alpha =(\mathbf{1},\mathbf{\sigma )}$ into
two classes : in one class, the matrix $M_3=\sigma _3$ commutes with $\Gamma
,$ we call it ''\textit{even matrix}''; the others $M_0=\mathbf{1,}$ $%
M_1=\sigma _1$ and $M_2=\sigma _2$ anti-commute with $\Gamma ,$ we call it ''%
\textit{odd matrices}''. Then one can expand $U(\mathbf{k})$ as $U(\mathbf{k}%
)=\sum\limits_\alpha u_\alpha (\mathbf{k})M_\alpha ,$ $\alpha =0,1,2,3$. We
can prove that $u_3(\mathbf{k})=u_3(-\mathbf{k})$ and $u_\alpha (\mathbf{k}%
)=-u_\alpha (-\mathbf{k})$, $\alpha =(0,1,2).$ Thus $u_\alpha (\mathbf{k})$
( $\alpha =(0,1,2)$ ) are odd functions of $k_x,k_y$ and are fixed to be
zero at momenta $(0,0)$, $(0,\pi )$, $(\pi ,0)$, $(\pi ,\pi )$ : $u_\alpha (%
\mathbf{k}=0)\equiv 0.$ Here $\mathbf{k}=0$ denote four special points in
momentum space $\left( 0,0\right) ,$ $(0,\pi )$, $(\pi ,0)$, $(\pi ,\pi )$ .

In general, $U(\mathbf{k})$ is written into
\[
U(\mathbf{k})=\left(
\begin{array}{ll}
\varepsilon _{\mathbf{k}} & \Delta _{1,\mathbf{k}}+i\Delta _{2,\mathbf{k}}
\\
\Delta _{1,\mathbf{k}}-i\Delta _{2,\mathbf{k}} & -\varepsilon _{\mathbf{k}}
\end{array}
\right)
\]
where $\varepsilon _{\mathbf{k}}$ is the even function of $k_x$ and $k_y$
and $\Delta _{1,\mathbf{k}},$ $\Delta _{2,\mathbf{k}}$ are the odd functions
of $k_x$ and $k_y$. At the points $\mathbf{k}>0,$ one has
\[
H(\mathbf{k}>0)_{\text{mean}}=\sum\limits_{\mathbf{k>}0}\left( \alpha _{%
\mathbf{k}}^{\dagger },\beta _{\mathbf{k}}^{\dagger }\right) \left(
\begin{array}{ll}
\varepsilon (\mathbf{k}) & 0 \\
0 & -\varepsilon (\mathbf{k})
\end{array}
\right) \left(
\begin{array}{l}
\alpha _{\mathbf{k}} \\
\beta _{\mathbf{k}}
\end{array}
\right)
\]
where $\alpha _{\mathbf{k}>0}$ and $\beta _{\mathbf{k}>0}=\alpha _{-\mathbf{k%
}}^{\dag }$ are (diagonalized) quasiparticle operators. Here $\mathbf{k}>0$
means that $k_y>0$ or $k_y=0,\ k_x>0$. At the four points $\left( 0,0\right)
,$ $(0,\pi )$, $(\pi ,0)$, $(\pi ,\pi ),$ the ``mean-field''\ Hamiltonian is
diagonalized into
\[
H(\mathbf{k}=\mathbf{0})_{\text{mean}}=\sum\limits_{\mathbf{k}=0}\varepsilon
(\mathbf{k})\psi _{\mathbf{k}}^{\dag }\psi _{\mathbf{k}}.
\]

The ground state is denoted by a projected $p$-wave BCS\ type wave-function
as
\begin{equation}
|g\rangle =|\Omega \rangle =P\prod_{\mathbf{k}}|u_{\mathbf{k}}|^{1/2}\exp {%
(\frac 12\sum_{\mathbf{k}}g_{\mathbf{k}}\psi _{\mathbf{k}}^{\dagger }\psi _{-%
\mathbf{k}}^{\dagger })}|0\rangle ,  \label{w}
\end{equation}
where $g_{\mathbf{k}}=v_{\mathbf{k}}/u_{\mathbf{k}}$ and $v_{\mathbf{k}%
}^2=\frac 12\left( 1-\frac{\varepsilon _{\mathbf{k}}}{E_{\mathbf{k}}}\right)
,\ u_{\mathbf{k}}^2=\frac 12\left( 1+\frac{\varepsilon _{\mathbf{k}}}{E_{%
\mathbf{k}}}\right) .$ Here $E_{\mathbf{k}}=\sqrt{\varepsilon _{\mathbf{k}%
}^2+(\Delta _{1,\mathbf{k}})^2+(\Delta _{2,\mathbf{k}})^2}$ is the energy
dispersion. $P$ is the projected operator which guarantees a constraint to
the Hilbert space with single fermion on each site. For the fermionic vacuum
we have ${\psi }_{\mathbf{k}}|0\rangle =0$ and all the negative energy
levels are filled by the fermions. Under these conditions, the ground state
of Eq.(\ref{w}) is not a p-wave superconducting state, instead it is an
insulator of topological order!

In the following part we will use the projected $p$-wave SC wave functions
to learn the nature of the topological orders, including the classification
of topological orders for spin models with translation invariance, the
degeneracy of the ground states and the prediction of new types of
topological orders.

\emph{Classification of topological orders }- Firstly, we classify the
topological orders on the lattice by the projected $p$-wave SC
wave-functions. Because there is no local order parameter to distinguish
different topological phases, we classify the topological phases by the
quantum phase transitions between them. Since the quantum states with
topological order are protected by the finite energy gap for excitations and
stable against any local perturbations, to break down a topological phase,
one needs to close the energy gap for the excitations\cite{fzx,kou}. That is
the \emph{unmovable gapless excitations} indicate a quantum phase transition
between different topological orders.

For the topological orders described by the projected $p$-wave SC
wave-functions, we find that $\mathbf{\varepsilon }(\mathbf{k}=0)=0$ at one
or more points of $\mathbf{k}=0$ define the quantum topological transitions
between different topological orders. For example, the topological order
describe by the projected $p$-wave SC wave-function with $\mathbf{%
\varepsilon }(\mathbf{k}=0)>0$ and that by the projected $p$-wave SC
wave-function with $\mathbf{\varepsilon }(\mathbf{k}=0)<0$ cannot be the
same one. The energy gap must vanish at $\mathbf{k}=0,$ corresponding to the
gapless critical states separating two topological phases. Then\emph{\ the
sighs of }$\varepsilon (\mathbf{k}=0)$\emph{\ become an additional structure
beyond projected spin group to distinguish topological phases}. There are
totally 16 different cases for the sighs of $\mathbf{\varepsilon }(\mathbf{k}%
=0)$ which characterize 16 different universal classes of topological
orders. We use the symbols ''$\mu \nu \lambda \xi "$ to denote them. $\mu ($%
or $\nu $, $\lambda $, $\xi $ $)=1$ denotes the case for $\mathbf{%
\varepsilon }(\mathbf{k}=(0,0))<0$ (or $\mathbf{\varepsilon }(\mathbf{k}%
=(\pi ,0))<0$, $\mathbf{\varepsilon }(\mathbf{k}=(0,\pi ))<0$, $\mathbf{%
\varepsilon }(\mathbf{k}=(\pi ,\pi ))<0$) and $\mu ($or $\nu $, $\lambda $, $%
\xi $ $)=0$ denotes the case $\mathbf{\varepsilon }(\mathbf{k}=(0,0))>0$ $($%
or $\mathbf{\varepsilon }(\mathbf{k}=(\pi ,0))<0,$ $\mathbf{\varepsilon }(%
\mathbf{k}=(0,\pi ))<0$, $\mathbf{\varepsilon }(\mathbf{k}=(\pi ,\pi ))<0).$

\begin{figure}[tbp]
\includegraphics[width=0.38\textwidth]{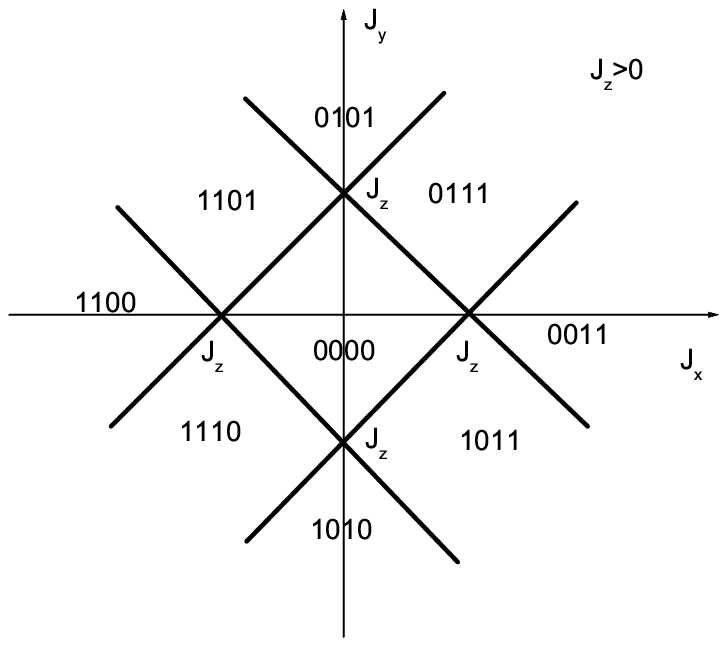}
\caption{Classification of topological orders by a p-wave SC wave function
for $J_z >0$. }
\label{Fig.1}
\end{figure}

\begin{figure}[tbp]
\includegraphics[width=0.38\textwidth]{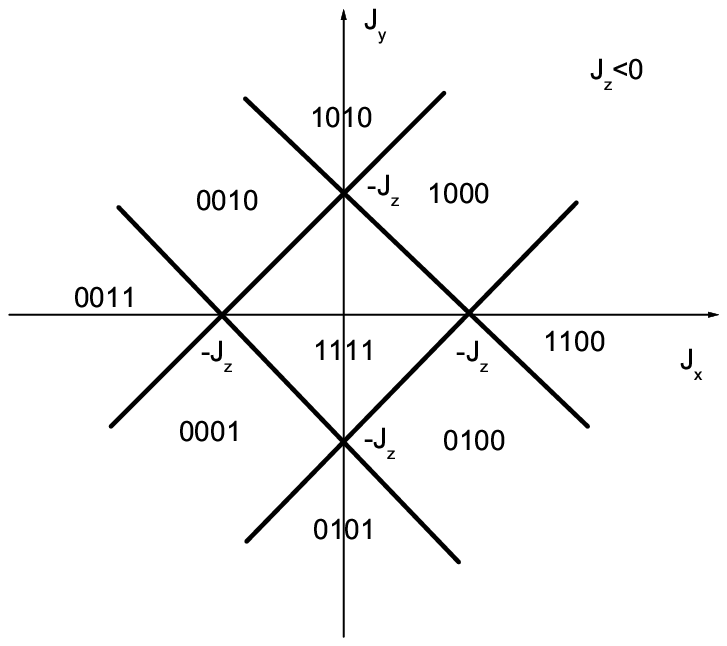}
\caption{Classification of topological orders by a p-wave SC wave function
for $J_z <0$. }
\label{Fig.2}
\end{figure}

\emph{Fermionic wave-functions for the topological orders} - let's give one
example for each class of topological order by the p-wave SC wave-functions.
A general mean field $p$-wave SC wave-function is written as

\[
U(\mathbf{k})=\left(
\begin{array}{ll}
\varepsilon _{\mathbf{k}} & \Delta _{1,\mathbf{k}}+i\Delta _{2,\mathbf{k}}
\\
\Delta _{1,\mathbf{k}}-i\Delta _{2,\mathbf{k}} & -\varepsilon _{\mathbf{k}}
\end{array}
\right) ,
\]
\begin{eqnarray}
\varepsilon _{\mathbf{k}} &=&J_z-J_x\cos k_x-J_y\cos k_y,  \nonumber \\
\Delta _{1,q} &=&J\sin k_x+J\sin k_y\text{, }\Delta _{2,\mathbf{k}}=0,
\end{eqnarray}
with $J\neq 0.$ The classification of the topological orders for above
wave-function is shown in the fig.1 and fig.2. From the two figures, one can
see that the topological orders of class $1001$ and class $0110$ cannot be
denoted by above wave functions. For $1001,$ instead, a given ansatz is $%
\varepsilon _{\mathbf{k}}=-J\cos (k_x+k_y),$ $\Delta _{1,\mathbf{k}%
}=J^{\prime }\sin k_x,$ $\Delta _{2,\mathbf{k}}=-J\sin (k_x+k_y).$ For $%
0110, $ a given ansatz is $\varepsilon _{\mathbf{k}}=J\cos (k_x+k_y),$ $%
\Delta _{1,\mathbf{k}}=J^{\prime }\sin k_x,$ $\Delta _{2,\mathbf{k}}=J\sin
(k_x+k_y).$

\emph{Topological degeneracies }- We calculate the topological degeneracies
for different classes of topological orders by using the p-wave SC
wave-functions. Now we assume $(u_{ij}^{(m,n)},$ $\eta
_{ij}^{(m,n)})=((-)^{ms_x(ij)}(-)^{ns_y(ij)}\bar{u}_{ij},$ $%
(-)^{ms_x(ij)}(-)^{ns_y(ij)}\bar{\eta}_{ij})$ to note four degenerate ground
states for the topological orders on a torus. Take $m,n=0,1$ as an example. $%
s_{x,y}(ij)$ have values $0$ or $1$, with $s_{x,y}(ij)=1$ if the link $ij$
crosses the $x$ or $y$ line and $s_{x,y}(ij)=0$ otherwise. However, some
ansatzs may be not permitted by the projections $P_r^e$ (or $P_r^o$)$.$ The
ground state $|\Psi _{\text{mean}}^{(u_{ij}^{(m,n)},\eta
_{ij}^{(m,n)})}\rangle $ with even (or odd) number particles is only allowed
under projection $P_r^e$ (or $P_r^o$),
\begin{equation}
|\Psi _{\text{spin}}\rangle =P_r^{e,o}|\Psi _{\text{mean}}^{(u_{ij}^{(m,n)},%
\eta _{ij}^{(m,n)})}\rangle =|\Psi _{\text{mean}}^{(u_{ij}^{(m,n)},\eta
_{ij}^{(m,n)})}\rangle .
\end{equation}

To obtain the topological degeneracy, one needs to deal with the projection
operators $P_r^e$ and $P_r^o$ carefully. Let's simplify the projection
operators firstly. We note that, for $\mathbf{k}>0$,
\begin{equation}
N_{\mathbf{k}>0}=\alpha _{\mathbf{k}>0}^{\dag }\alpha _{\mathbf{k}>0}-\beta
_{\mathbf{k}>0}^{\dag }\beta _{\mathbf{k}>0}+1.
\end{equation}
For the mean-field ansatz defined above, there always exists a negative
energy level\ for fermions at the point $\mathbf{k}>0$. For the energy of $%
\beta $\ band $-\varepsilon (\mathbf{k})$ is negative, $\varepsilon (\mathbf{%
k})>0,$ $\beta $\ band is occupied and $\alpha $\ band is empty, we have $N_{%
\mathbf{k}>0}=0 $. On the other hand, for the energy of $\alpha $\ band $%
\varepsilon (\mathbf{k})$ is negative, $\varepsilon (\mathbf{k})<0,$ $\alpha
$\ band is occupied and $\beta $\ band is empty, we have $N_{\mathbf{k}>0}=2$%
. As a result, the total number of $\psi $\ fermions on every point of $%
\mathbf{k}>0 $, $\sum\limits_{\mathbf{k}>0}N_{\mathbf{k}}$,\ is even. Then $%
(-1)^{\hat{N}}$ is reduced into $(-1)^{\hat{N}_0}$ with $\hat{N}%
=\sum\limits_{\mathbf{k}>0}N_{\mathbf{k}}+\hat{N}_0.$ Here $\hat{N}%
_0=\sum\limits_{\mathbf{k}=0}\psi _{\mathbf{k}}^{\dag }\psi _{\mathbf{k}}$
is the total fermion number on the four special points $\mathbf{k}=0$.

So to determine if the mean-field ground state contain even or odd number of
$\psi $ fermions, we only need to examine the occupation on the four special
points: $(0,0)$, $(0,\pi )$, $(\pi ,0)$, $(\pi ,\pi )$ and calculate the
total fermion number on them$.$ The projection operators now are reduced
into $P_r^e=\frac{1+(-1)^{\hat{N}_0}}2$ and $P_r^o=\frac{1-(-1)^{\hat{N}_0}}%
2 $. For a given point $\mathbf{k}=0,$ the energy for $\psi $ is $\mathbf{%
\varepsilon }(\mathbf{k}=0).$ Then if $\mathbf{\varepsilon }(\mathbf{k}%
=0)<0, $ one $\psi $ particle occupies energy level at the point $(0,0)$ (or
$(0,\pi ),$ $(\pi ,0),$ $(\pi ,\pi )$)$;$ if $\mathbf{\varepsilon }(\mathbf{k%
}=0)>0$, zero $\psi $ particle occupies the energy level at $(0,0)$ (or $%
(0,\pi ),$ $(\pi ,0),$ $(\pi ,\pi )$). Finally we obtain the projection
operators as
\[
\frac{1\pm (-1)^{\hat{N}_0}}2=\frac{1\pm (-1)^{\sum\limits_{\mathbf{k}%
=0}\zeta (\varepsilon (\mathbf{k}))}}2
\]
where the function $\zeta \left( x\right) $ denotes $\zeta \left( x\right)
=1,$ for $x<0$ and $\zeta \left( x\right) =0,$ for $x>0$. One can check the
occupation number at $\mathbf{k}=0$ easily to know if a given ansatz is
valid, Using this method, we obtain the topological degeneracy on different
lattices for the $16$ topological orders. The results are given in the
following tables in detail :

\begin{widetext}

\[
\begin{array}{lllllllllllllllll}
& 1111 & 1110 & 1101 & 1011 & 0111 & 1100 & 1010 & 1001 & 0110 &
0011 & 0101
& 1000 & 0100 & 0010 & 0001 & 0000 \\
(e*e) & 4 & 3 & 3 & 3 & 3 & 4 & 4 & 4 & 4 & 4 & 4 & 3 & 3 & 3 & 3 & 4 \\
(e*o) & 4 & 3 & 3 & 3 & 3 & 4 & 2 & 2 & 2 & 4 & 2 & 3 & 3 & 3 & 3 & 4 \\
(o*e) & 4 & 3 & 3 & 3 & 3 & 2 & 4 & 2 & 2 & 2 & 4 & 3 & 3 & 3 & 3 & 4 \\
(o*o) & \_ & 3 & 3 & 3 & 3 & 2 & 2 & 2 & 2 & 2 & 2 & 1 & 1 & 1 & 1 &
4
\end{array}
\]

\end{widetext}

From the results, we can see that for the $16$ topological orders, there are
$8$ Z2 topological orders ( $1111,$ $1100,$ $1010,$ $1001,$ $0101,$ $0011,$ $%
0110,$ $0000$) and $8$ non-Abelian topological orders ($1000,$ $0100,$ $%
0010, $ $0001,$ $1110,$ $1101,$ $1011,$ $0111$).

The Z2 topological orders are described by Z2 projective symmetry groups.
Such Z2 topological orders always have $4$ degenerate ground states on an
even-by-even lattice with periodic boundary condition. And there always
exists three types of quasiparticles: Z2 charge, Z2 vortex, and fermions,
respectively\cite{wen2}. And the fermions can be regarded as bound states of
a Z2 charge and a Z2 vortex.

On the other hand, we have two classes of $8$ types the non-Abelian
topological orders\cite{comment} : on an even-by-even lattice the
topological degeneracy is always $3$. However, on an odd-by-odd lattice, the
degeneracy is $3$ for first class $4$ non-Abelian topological orders and $1$
for the other class $4$ topological orders. All the $8$ non-Abelian
topological orders have the two types of quasiparticles: non-Abelian anyons
and fermions.

Our results are consistent to those by Read and Green \cite{rg}. They have
pointed out there are two kinds of $p_x+ip_y$-wave fermion paired state :
one is in the weak pairing phase as an example of topological order with
non-Abelian anyon \cite{rg}, the other in the strong pairing phase as a
topological order with Abelian anyon. However, our results show more rich
physics properties for the $p_x+ip_y$-wave fermion paired states on lattice
beyond the picture of weak and strong pairing phases.

\emph{Unification of the topological orders for exact solved spin models }-
Finally we can identity the classes of topological orders for the exact
solved spin models by our classification and show their topological
properties.

\label{section-fermionization}For the Kitaev model on a hexagonal lattice
\cite{k2}, the original spin model is mapped to a model of $p$ -wave BCS
model on a square lattice by a Jordan-Wigner transformation with \cite{cn}
\begin{eqnarray}
\varepsilon _{\mathbf{k}} &=&\left| J_z\right| -\left| J_x\right| \cos
k_x-\left| J_y\right| \cos k_y,  \nonumber \\
\Delta _{1,q} &=&\left| J_x\right| \sin k_x+\left| J_y\right| \sin k_y\text{%
, }\Delta _{2,\mathbf{k}}=0.  \label{eD}
\end{eqnarray}
From the relationship between $\left| J_x\right| ,$ $\left| J_y\right| $, $%
\left| J_z\right| ,$ we found that the phase $|J_z|>|J_x|+|J_y|$ as $0000$
type Z2 topological order; $|J_y|>|J_x|+|J_z|$ as $1010$ type Z2 topological
order; $|J_x|>|J_y|+|J_z|$ as $1100$ type Z2 topological order. The ground
state degenerate is always 4 on an even-by-even lattice (on a torus).
However, on other lattices (even-by-odd, odd-by-even, odd-by-odd), the
ground state degenerate is different : for 0000 type, it is 4; for 1010; it
is 4 on an odd-by-even lattice, but 2 on an even-by-odd and odd-by-odd
lattice; for 1100; it is 4 on an even-by-odd lattice, but 2 on an
odd-by-even and odd-by-odd lattice.

For the Kitaev model on a honeycomb lattice with minimal three- and
four-spin terms and a T-symmetry breaking external magnetic field, the
corresponding $p$-wave SC state is given as \cite{yu,ki,zhou}
\begin{eqnarray}
\varepsilon _{\mathbf{k}} &=&\left| J_z\right| -\left| \tilde{J}_x\right|
\cos k_x-\left| \tilde{J}_y\right| \cos k_y,  \nonumber \\
\Delta _{1,\mathbf{k}} &=&\Delta _{1x}\sin k_x+\Delta _{1y}\sin k_y,
\nonumber \\
\Delta _{2,\mathbf{k}} &=&\Delta _{2x}\sin k_x+\Delta _{2y}\sin k_y
\end{eqnarray}
with $|J_z|<|\tilde{J}_x|+|\tilde{J}_y|,$ $|J_z|>|\tilde{J}_x|-|\tilde{J}_y|$%
, $|J_z|>-|\tilde{J}_x|+|\tilde{J}_y|$. It is described by the $1000$ type
Non-Abelian topological order\cite{k2,lee}. The topological degeneracy is $3,
$ $3,$ $3,$ $1$ for even-by-even, even-by-odd, odd-by-even, odd-by-odd
lattices, respectively.

For the Wen's plaquette model, a ``mean-field'' ansatz $p$-wave BCS states
becomes\cite{wen,wen1}
\begin{equation}
H_{mean}=\sum_{\langle ij\rangle }\left( \psi _{I,i}^{\dag }u_{ij}^{IJ}\psi
_{J,j}+\psi _{I,i}^{\dag }\eta _{ij}^{IJ}\psi _{J,j}^{\dag }+h.c.\right)
\end{equation}
where $I,J=1,2$. The ground state for $g<0$ is a topological state with $%
-\eta _{i,i+x}=u_{i,i+x}=1+\sigma ^3$ and $-\eta _{i,i+y}=u_{i,i+y}=1-\sigma
^3$. The quantum wave-function can be regarded as two-flavor spinless
projected $p$-wave SC. Then the fermion number on each lattice and the total
fermion number must be even. Using the same method, one can also calculate
the ground state degeneracy which is $4,$ $2,$ $2,$ $2$ for even-by-even,
even-by-odd, odd-by-even, odd-by-odd lattices, respectively. It is the $0110$
type of the topological order.

In addition, it is known that the topological orders of the toric-code model
and the Wen-plaquette model are the the same ones\cite{k1}. So one can use
the same $p$-wave SC wave-function in Eq.(8) to describe the toric code
model.

In conclusion, we develop a sysmatical theory for topological orders by the
(projected) $p$-wave SC wave-functions. Based on the theory we unify
different topological ordered states for the exact solved spin models and
obtain their topological properties. From our classification in this paper,
the topological orders in spin models are the same one and at least five
different classes of topological orders have been found. That is one cannot
change the topological state for the Wen-plaquette model into that of the
Kitaev model without a quantum phase transition. And we can predict that at
the quantum phase transition between them the massless excitations should
have zero energy at $\mathbf{k}=0$. In the end, we can also give a
prediction that there are totally $11$ unknown classes topological orders,
including four classes of Z2 topological order ( $1111,$ $1001$, $0011$, $%
0101$ ) and $7$ classes of non-Abelian topological orders. All the new
classes of topological orders need to be explored in the future.

The email for Su-Peng Kou is spkou@bnu.edu.cn. S.P. Kou acknowledges that
this research is supported by NFSC Grant no. 10574014.


\begin{thebibliography}{99}
\bibitem{comment}  For a non-Abelian topological phase, there exists another
topological invariable - the Chern number defined by $v=\frac 1{2\pi i}\int
Tr(U(\mathbf{k})dU(\mathbf{k})\wedge U(\mathbf{k}))$ with $U(\mathbf{k}%
)=\sum\limits_\alpha u_\alpha (\mathbf{k})M_\alpha $ \cite{rg, k2}. In our
classifiaction, we consider a fixed Chern number.
\end{thebibliography}
\end{document}